\documentclass[
preprint,
aps,
showpacs, showkeys, longbibliography, secnumarabic, floats
]{revtex4}%
\usepackage[figurename=Figure]{caption}
\usepackage{amsfonts}
\usepackage{amsmath}
\usepackage{amssymb}
\usepackage{graphicx}
\usepackage[papersize={25cm,35cm}]{geometry}
\usepackage[colorlinks=true,linkcolor=blue,pagecolor=blue,citecolor=green,menucolor=blue]{hyperref}
\usepackage{url}%
\providecommand{\U}[1]{\protect\rule{.1in}{.1in}}

\DeclareCaptionLabelSeparator{space}{ }
\begin{document}
\preprint{cond-mat.mtrl-sci}
\title[Nanoscale measurement of the Power Spectral Density of surface roughness: how to solve a difficult experimental challenge]{Nanoscale measurement of the Power Spectral Density of surface roughness: how to solve a difficult experimental challenge}
\author{Juan Francisco Gonz\'{a}lez Mart\'{i}nez}
\email{jfgm@um.es}
\author{In\'{e}s Nieto Carvajal}
\email{inc2@um.es}
\author{Jos\'{e} Abad}
\email{jabad@um.es}
\author{Jaime Colchero Paetz}
\email{colchero@um.es}
\affiliation{Universidad de Murcia}
\date{\today}

\begin{abstract}

In the present work we show that the correct determination of
surface morphology using Scanning Force Microscopy (SFM) imaging and
Power Spectral Density (PSD) analysis of the surface roughness is an
extremely demanding task that is easily affected by experimental
parameters such as scan speed and feedback parameters. We present
examples were the measured topography data is significantly
influenced by the feedback response of the SFM system and the PSD
curves calculated from this experimental data do not correspond to
that of the true topography. Instead, either features are ``lost''
due to low pass filtering or features are ``created'' due to
oscillation of the feedback loop. In order to overcome these serious
problems we show that the interaction signal (error signal) can be
used not only to quantitatively control but also to significantly
improve the quality of the topography raw data used for the PSD
analysis. In particular, the calibrated error signal image can be
used in combination with the topography image in order to obtain a
correct representation of surface morphology (``true'' topographic
image). From this ``true'' topographic image a faithful
determination of the PSD of surface morphology is possible. The
corresponding PSD curve is not affected by the fine-tuning of
feedback parameters, and allows for much faster image acquisition
speeds without loss information in the PSD curve.

\end{abstract}

\pacs{07.79.Lh}
\keywords{Power Spectrum Density, PSD, error signal, Scanning Force Microscopy, surface roughness, feedback.}\maketitle

\section{Introduction}

The nanoscale surface morphology determines a wealth of phenomena
which are important for fundamental science as well as for
technological applications \cite{Bhushan,Persson2005,Zap}. As is
well known, surface roughness is a basic parameter in tribology
\cite{Tribo1,Tribo2}, adhesion phenomena \cite{Adh,Persson2005}, the
internal 3--D morphology of nanostructural functional materials
\cite{het}, wetting properties of surfaces
\cite{Lotus,wetting1,wetting2,Watson}, optical reflectivity
\cite{optical1,optical2} as well as a wealth of biological processes
\cite{Lotus,Bio}. A precise and reproducible measurement of surface
roughness is therefore a key issue for basic science as well as for
engineering applications \cite{Adif}. Accordingly, important efforts
have been undertaken in this field ranging from the development of
suitable instruments to the normalisation and traceability of length
measurements. Traditionally, surface roughness has been measured
using profilometers \cite{Adif,Thomas1975205}, although optical
instruments have proven to be very powerful tools as well
\cite{Optics,OpticalProfilometer}. With the invention of the Scanning Tunneling
Microscope \cite{Binnig} and later, the Scanning Force Microscope
(SFM) \cite{SFM}, it has been possible to determine the morphology
of surfaces down to the nanometer and even the atomic scale.
Accordingly scanning tunneling microscope and scanning force
microscope have been widely used for nanoscale roughness
characterization \cite{Baro1,Baro2,nanorou}.

The morphology of surfaces can be described using a variety of
parameters, the Root Mean Square (RMS) roughness is surely the most
common one \cite{RMS1,RMS2}. In addition, other parameters such as
Skewness and Kurtosis can also be used to characterize a surface.
Unfortunately, these parameters do not describe the morphology of
surfaces in a sufficiently accurate way. This can be understood
already on very simple arguments: if we assume that a surface has
been discretised using $n\times n$ image points, the whole
information content of the surface is reduced to one single value if
only the RMS value of the surface is measured. Surfaces with very
different morphology -- and thus very different behavior with regard
to tribology or adhesion -- may have the same RMS value of surface
roughness. Even worse, it can be shown that for many interesting
surfaces the RMS roughness depends on the length scale used for the
measurement; that is, as the size of an image is increased, the
measured RMS also increases. The RMS-value of surface roughness is
therefore not a scale invariant quantity. The precise description of
surface morphology therefore calls for more sophisticated tools. As
discussed in more detail elsewhere, the Power Spectral Density (PSD)
of surface roughness is such a tool \cite{Persson2005,Fang}. PSD in
combination with SFM is an invaluable tool in nanoscale science that
should be further developed to really exploit all its possibilities
\cite{Baro2,Ruf,het,Quim1,ZhouAFMPSD}.

Essentially the PSD describes the mean surface roughness at each length scale in a given image. Typically an
image with $n\times n$ points results in a PSD curve with $n/2$ points.
Evidently, even though the reduction of information content is quite high, the
reduction is much less as compared to the case where only the RMS is computed.
Interestingly, for many surface the roughness varies in a well defined way as
the length scale is varied (so-called self afine surfaces). In this case the
PSD curve is particularly simple since the relation $\log\left[
\text{PSD}(\lambda)\right]  $ vs $\log(\lambda)$, where $\lambda$ is the
wavelength of the surface roughness, is
linear: $\log\left[\text{PSD}%
(\lambda)\right]  $ vs $\log(\lambda)=s\cdot\log(\lambda)+b$ (see, for example \cite{Persson2005}).

From a theoretical point of view, description of surfaces using PSD
curves is a very powerful tool and is the basics for modern
approaches relating tribology and adhesion phenomena to microscopic
and nanoscale properties of surfaces \cite{Persson2005}.
Unfortunately, experimental determination of nanoscale PSD is quite
demanding. Due to the compressing properties of the logarithm a
large number of image points are needed to have a significant amount
of data points for the horizontal axis ($\log(\lambda)$). Moreover,
since the relevant magnitude of the the vertical axis is also
logarithmic ($\log\left[ \text{PSD}(\lambda)\right]  $) a large
dynamic range for the height measurement is also required, that is,
very large as well as very small height differences have to be
acquired equally well. Data adquisition in Scanning Probe Microscopy
is sequential and thus inherently slow as compared to other imaging
techniques which are generally based on parallel processing (as in
most optical microscopes). As is well known, Scanning Probe
techniques are based upon a very short range of interaction between
the sample to be analyzed and a sharp tip used as probe. In order to
obtain surface morphology, the tip is scanned over the sample while
a feedback loop is used to keep tip-sample interaction at a constant
value. As the tip moves over the sample, this feedback adjusts the
(absolute) height of the tip in order to compensate for height
variations of the sample surface. Correct adjustment of the feedback
parameters is fundamental for the acquisition of good topography raw
data: slow feedback will result in ``smoothing'' of topography data,
which would effectively imply low pass filtering and a loss of high
frequency roughness, while fast feedback may induce feedback
oscillation, creating artificial high frequency roughness not
present in the true topography of the sample. Although these
feedback issues are always present in SPM experiments and are
generally solved in an intuitive way, they are specially relevant if
the PSD of surface roughness is determined. Indeed, for its faithful
determination each image point of the topography raw data has to be
acquired faithfully. For rough surfaces -- those for which a PSD
analysis is particularly interesting-- this is a very difficult
experimental challenge: feedback parameters have to be adjusted for
fast feedback but avoiding oscillations and the imaging speed has to
be chosen to allow correct settling of the feedback at each image
point (what is ``correct'' in this context?). In principle, large
enough acquisition times (low acquisition speeds) would allow
correct measurement of topographic data; however, for large images
(more than $10^{3}\times10^{3}=$ a million data points!) this may
result in unpractical acquisition times for a single image (up to
days!). Moreover, large acquisition times would induce additional
problems due to low frequency noise and drift, which would also
distort the real topography of the sample \cite{InesPsd1D}.

In the present work we will address in detail these issues. We will show that
the interaction signal (error signal) can be used not only to quantitatively
control but also to significantly improve the quality of the topography raw
data used for the PSD analysis. Firstly, we will investigate the effect of
feedback response on the determination of the PSD. We will find that,
unfortunately, the PSD strongly depends on the setting of the feedback loop
(proportional and feedback parameters) as well as on the imaging speed.
Secondly, we will show that if the error signal used for the feedback is
appropriately calibrated, then the correct determination of the PSD can be
significantly improved. In particular, the calibrated error signal can then be
used in combination with the topography image for a faithful determination of
the surface morphology from which the correct PSD measurement is obtained.
Moreover, the corresponding PSD curve is much less affected by a correct
fine-tuning of feedback parameters, and allows for much faster image
acquisition speeds without loss of information in the PSD curve.

\section{Experimental and Data Processing}

Experiments were performed using a NanoTec SFM system composed of
SFM head, high voltage controller and PLL/dynamic measurement board
\cite{Nanotec}. In this kind of experiments we use sharpened tips
with a force constant of 2 N/m and a resonance frequency around
70kHz \cite{Olympus}. In order to obtain maximum stability of the
mechanical set-up, the microscope was kept working overnight before
the relevant measurements were performed. For the experiments
discussed in the present work we operate the Scanning Force
Microscope in dynamic mode using the oscillation amplitude as
feedback channel (AM-DSFM). Relatively large oscillation amplitudes
(50-80 nm peak-peak) and significant reduction of oscillation
amplitude are used for feedback (setpoint of 0.5-0.75 a$_{\text{free}}$
with a$_{\text{free}}$ the free oscillation frequency) are used for
feedback. We note that, contrary to typical AM-DSFM operation in air, a
phase locked loop within the dynamic measurement board is enabled to
keep the tip sample system always at resonance.

As will be discussed in detail elsewhere, we use the thermal noise
to precisely calibrate the oscillation amplitude \cite{JuanFran1}. We have recently
shown how the amplitude signal, and in particular the thermal noise
of the amplitude signal, is processed by the electronics of a
dynamic unit (usually a lock-in scheme together with a phase locked
loop to follow the resonance frequency)\cite{Ruido}.

In order to calibrate de amplitude signal, the dynamic unit is used to
demodulate the thermal noise signal which is shifted to $\Delta \nu_{\text{A}}=\nu-\nu_{\text{ref}}$ and $\Delta \nu_{\Sigma}=\nu+\nu_{\text{ref}}$ in the
``amplitude'' and ``phase'' output of a typical dynamic unit. The corresponding spectrum can then be
acquired either with a signal analyser, or from the Fourier
transform of the time domain signal of the oscillation amplitude.
From the equipartition theorem we obtain $kT/2=c<a^2>/2$ , where $kT$ is the
thermal energy, $c$ the force constant of the cantilever and $<a^2>$
the square of the rms amplitude signal and from this relation a calibration factor
for the oscillation amplitude is obtained. We note that a very low
noise level of the detection electronics is needed for this
technique to work, since it essentially assumes that all the noise
measured is thermal noise. Finally, we also note that a precise
calibration of oscillation amplitude (better than 5\%) is essential
for the work discussed here.

The PSD of an image is usually calculated from the 2--dimensional
Fourier Transform of a topographic image by angle averaging the
Fourier transform in all directions \cite{Persson2005}. Another
possibility is to compute the 1--dimensional PSD of each (horizontal)
line of an image and then average the power spectrums obtained from
all lines of the image. The latter approach is used in the present
work and will be discussed in detail elsewhere \cite{InesPsd1D}. PSD
curves have been computed using a specifically programmed
Mathematica$^\circledR$ code or directly within the WSxM$^\circledR$
software \cite{WSxM,wetting2}.

\section{Effect of Feedback response on the determination of surface
morphology and Power Spectral Density of surface roughness}

To illustrate the problem of feedback response on the determination
of the PSD curve \textbf{figure \ref{Fig1}} shows a series of
topographic images of a glass cover slide. Images were acquired with
the same imaging speed (1 line/s) but different proportional/
integral (P/I) feedback parameters. The cover slide has been
carefully cleaned in order to remove any contaminations from the
surface. We note in this context that already a small number of
contamination particles (``nanoscale dust'') on the glass surface
will change the PSD curve obtained from an experimental image, and
thus affect the statistical properties of the measured surface
morphology. As discussed in detail elsewhere, a typical glass
surface should have a self-affine structure with a fractal dimension
$D_{f}=1.5$ \cite{Persson2005}. Such a fractal surface was
considered particularly appropriate for the present study, since it
will have surface roughness at all length scales. We recall that an
ideal self-affine surface will show a linear relation of the surface
roughness in a log-log plot: $\log\left[  \text{PSD}(\lambda)\right]
$ vs $\log(\lambda)=s\cdot \log(\lambda)+b$. For the case of the PSD
curves shown (averages of the PSD curves for each line),
a fractal dimension $D_{f}=1.5$ should result in a
slope $s=2D_{f}-5=-2$ \cite{Persson2005}. In addition, a
self-affine surface should present a characteristic disordered
appearance having large as well as small scale structures. More
precisely, it should have larger (=higher) ``large scale
structures'' and smaller (=lower) ``small scale structures''. In our
experiments, this ``cloudy'' appearance is best recognised in the
insets of the larger topographic images.

\textbf{Figure \ref{Fig2}} shows all PSD curves calculated from the
different topographic images shown in \textbf{figure \ref{Fig1}}.
In addition, a master curve is shown for comparison, which gives the ``true''
PSD, to be discussed in detail below. As expected from the arguments
discussed in the introduction, images acquired with different
setpoints of the feedback loop indeed result in quite different PSD
curves. The PSD curve obtained from the image with the highest P/I
values of the feedback loop gives the highest values for the surface
roughness. This PSD curve shows three clear peaks, which are only
recognised in the PSD curve but not in the corresponding topographic image,
where they are essentially imperceptible even in the zooms of the large scale
image. As the P/I values are decreased, the measured surface
roughness also decreases. The measured curves do not vary in a
simple way since the surface roughness is ``lost'' differently for
large and small length scales. In particular, the three peaks observed
in the ``fasted'' image, strongly decrease when the P/I values are
decreased. Moreover, the ``loss'' of surface roughness affects the
overall shape of the curve, and in particular its slope, from which
the fractal dimension is determined. Finally, we note that the
``cleanest'' curves with a relatively linear shape are obtained for
the lowest values of the P/I parameters. Intuitively we would expect
the middle curves to be the better ones because low frequency
components are not lost (too much?), and no high frequency
components are \textquotedblleft produced\textquotedblright\ due to
feedback oscillations. However, how can we assure that this argument
is correct? Can we define precise criteria in order to choose the
correct PSD-curve? To address this issue, in the next section we
will present a simple model in order to relate the measured
topography with the true topography and the measured tip-sample
interaction. Nevertheless, and in order to stress the importance of
this issue, \textbf{figure \ref{Fig2}} shows a ``master curve'' representing the true PSD curve of the glass
cover slide. This ``master curve'' will be discussed in detail below. We note that --quite disturbingly-- none of the PSD
curves obtained from the measured images coincides with the correct
(!) ``master curve''.

\section{Simple modeling of the imaging adquisition process}

The key idea in the present section is that topography and error signal should
be complementary if appropriate measuring units are utilized. Essentially,
this is the key point of the present work. Note that if feedback is slow so
that small scale features are filtered in the topography image, these features
will appear in the signal that is used to maintain a constant tip-sample
interaction (error signal). On the contrary, if feedback were perfect --which
is unphysical-- the interaction signal would be constant and all information
would be in the topographic image. Finally, if the feedback oscillates, this
oscillation should be visible in both, in the topographic and in the error
signal image. In order to analyze this point further, we recall that
topography images are acquired by maintaining constant the interaction between
tip and sample as the sample is scanned; that is, the feedback should fulfill
the mathematical condition%
\begin{equation}
I(x,y,z(x,y))=I_{\text{set}} \label{implicit}%
\end{equation}
where $I_{\text{set}}$ is the setpoint for the feedback, and $z(x,y)$ is the surface
profile followed by the tip. For a given interaction field $I(x,y,z)$, the SPM
system therefore \textquotedblleft solves\textquotedblright\ the implicit
equation (\ref{implicit}) for the surface profile $z(x,y)$. In most cases the
force field is complicated and highly non--linear and may even depend on the
chemistry of the sample \cite{HighLow}. Then the surface profile depends in a non-trivial way
on the set point chosen for image acquisition: $z(x,y)=z(x,y,I_{\text{set}}%
)$. To keep the present analysis simple, we will
assume that effects due to non-linearity and surface chemistry are
not relevant for the experiments discussed here, that is, we will
asume that for a (reasonable) variety of setpoints the measured
profile does not depend on the setpoint chosen for the feedback
loop.

For a real, non-ideal feedback loop the surface profile $z_{\text{fb}}(x(t))$
followed by the tip of the SFM system will deviate from the true surface by
some error profile $\delta z_{\text{err}}(x(t))$:%
\begin{equation}
z_{\text{fb}}(x(t))=z_{\text{true}}(x(t))+\delta z_{\text{err}}(x(t))
\end{equation}
where we have assumed that only the fast scanning direction $x$ is relevant
for the present discussion and have thus omitted the slow scan direction $y$, because for the $y$ direction the feedback loop has sufficient time to settle.
In order to keep the notation simple, in what follows we will also omit the
time dependence of the signals. Note, however, that this dependence is quite
important since a faster scan $x_{\text{fast}}(t)$ will imply more error signal and a
different surface profile $z_{\text{fb}}(x_{\text{fast}}(t))$.

For a given surface profile $z_{\text{fb}}(x)$ the interaction signal which is measured will be
\begin{align}
I_{\text{meas}}(x,z_{\text{fb}}(x)) &  =I(x,z_{\text{true}}(x)+\delta
z_{\text{eff}}\left(  x\right)  )=\nonumber\\
&  =I(x,z_{\text{true}}(x))+\frac{\partial I}{\partial z}(x,z_{\text{true}%
}(x))~\delta z_{\text{eff}}+\ldots\label{ErrorSignal}\\
&  \simeq I_{\text{set}}+\Delta I_{\text{err}}(x,z_{\text{true}}(x))\nonumber
\end{align}
where we have kept only linear terms in the expansion of the interaction field
$I(x,z)$. In a real experiment, the measured interaction signal
$I_{\text{meas}}$ therefore deviates from the chosen setpoint I$_{set}$ by the
error signal $\Delta I_{\text{err}}$\ defined above. This deviation is caused
by a finite feedback response which results in a time lag between the
\textquotedblleft ideal\textquotedblright\ height of the tip ($z_{\text{true}%
}(x)$) and the height which is reached by the feedback loop ($z_{\text{fb}%
}(x)$). Since the tip moves over the surface, by the time the feedback would
have settled to the correct height the tip is at a new (lateral) position
$x+\delta x$ where the tip-sample height $z_{\text{true}}(x+\delta x)$ in
general will be different and the height $z_{\text{fb}}(x+\delta x)$ found by
the feedback loop is, again, not correct. Therefore, a SPM system does not
measure the real topography $z_{\text{true}}(x)$, but some other profile
$z_{\text{fb}}(x)$. As discussed previously, the amount of error will depend
on the scan speed. For linear systems the error is expected to be proportional
to the scan speed. An important consequence of relation \eqref{ErrorSignal} is
that the error profile $\delta z_{\text{eff}}\left(  x\right)  $ can be
obtained if the ``calibration factor''%
\ $\partial I/\partial z$ is known:
\begin{equation}
\delta z_{\text{err}}=\Delta I_{\text{err}}(x))\Big/\frac{\partial I}{\partial
z}(x,z_{\text{true}}(x))
\end{equation}
With this error profile, the true topography can be obtained directly from the
measured topography $z_{\text{fb}}$ and the error signal $\Delta
I_{\text{err}}$:%
\begin{align}
z_{\text{true}} &  =z_{\text{fb}}-\delta z_{\text{err}}=\nonumber\\
&  =z_{\text{fb}}(x,I_{\text{set}})-\Delta I_{\text{err}}(x)\Big/{\frac{\partial
I}{\partial z}(x,z_{\text{true}}(x))}%
\end{align}
Unfortunately the slope of the interaction is a quantity which is not
easy to determine. Moreover, as discussed above, generally the interaction is
non--linear, therefore its slope will vary with tip--sample distance and thus
with the set point chosen for constant interaction images. There are, however,
two important SFM modes where the interaction signal can be calibrated
appropriately and where the error signal depends linearly on tip--sample
distance for a suitable range of tip--sample interaction: the normal force
signal in the case of contact mode SFM and the oscillation amplitude in the
case of the so called AM-DSFM mode. Moreover, in these SFM modes the
interaction signal can be calibrated appropriately so that the error signal
can be specified directly in length units (nanometers); that is, the
interaction signal is then normalised so that conversion factor $\partial
I/\partial z$ is unity, thus $\Delta I_{\text{err}}=\delta z_{\text{err}}$. For
contact mode SFM the interaction signal is then essentially the (static
deflection) of the cantilever, while in case of AM-DSFM mode the interaction
signal is the oscillation amplitude of the cantilever. In these two SFM modes
the true topography is directly obtained by simple substraction of the
topography and error signal data:%
\begin{equation}
z_{\text{true}}=z_{\text{fb}}-\Delta I_{\text{err}}\label{ZtrueSumOrDiff}%
\end{equation}
Therefore, from the topographic and the error image the true topography of a
sample can be obtained, which is the raw data for the precise determination of
the PSD curve. Evidently, the true topography does not depend on the
particular set of parameters used for the feedback loop. When the
experimentally measured topography and the error signals are considered
uncorrelated entities, both will depend in a very strong way on these
parameters. Experimentally -- as shown in the next section -- only the combination
of both, topographic and error signal, is therefore a quantity (i.e. the true
topography) that does not depend on a particular set of parameters used for
data acquisition (proportional/integral parameters, scan speed, etc.).

\section{Determination of true surface morphology using topographic and error
signal data}

To illustrate the issues discussed in the preceding section,
\textbf{figure \ref{Fig3}} shows a series of images of a
Platinum film evaporated onto a Silicon surface. Data was acquired
in the constant amplitude Dynamic Scanning Force Microscopy mode
(also called AM-DSFM). Such a film presents a grain like structure,
with a typical lateral grain size of about 50 nm. These Platinum grains
can be clearly resolved in the enlarged areas of most images.
Topography as well as the corresponding amplitude (=error signal)
image are presented. As discussed in the experimental section, the
amplitude images have been carefully calibrated in order to
determine the precise oscillation amplitude. This allows to show the
amplitude images in length units (nm). Therefore all data
-topography as well as amplitude images- can be shown with the same
units and the same scale, in this case 5 nm. Images have been
acquired at a scanning rate of 1 line/s, but with different
feedback parameters: the images in the first column have been
acquired with the ``fastest'' feedback (high values for the
proportional/integral parameters) while those in the last column
have been acquired with the ``slowest'' feedback parameters.
Correspondingly, the Platinum grains are visualised in the
topography (top row) when the feedback is ``fast'', and in the
amplitude (middle row) when the feedback is ``slow''. Note that the
first topography image is essentially equivalent to the last
amplitude image, that is, visually the contrast of the grains is the
same. This proves that the calibration of the amplitude image is
correct, otherwise the height of the grains would appear different
in the topography and the amplitude image (recall that the grey
scale of the images is the same for all images).

In addition to the topography and error signal images, for each data set PSD
curves have been computed for both signals individually as well as for the
difference $z_{d}(x,y)=z_{\text{fb}}(x,y)-a(x,y)$. As expected from the
discussion in the previous section, PSD curves obtained from the individual topography and error
signal images strongly depend on the feedback parameters. The PSD curves of
the difference image give always, within our experimental error, the same ``master PSD curve''.

Before further discussing the different PSD curves we note that contrary to
the case of the glass slide the (``good'') PSD curves of the Platinum grains
are not linear, instead they saturate for low spacial frequencies (large
scales). In order to understand this behaviour, we propose a simple model for
the morphology of this surface: a disordered arrangement of individual
gaussian grains with a fixed height $h_{0}$ and a fixed lateral dimension
$w_{0}$,%
\[
z_{Pt}\left(  x,y\right)  =h_{0\;}e^{-(x^{2}+y^{2})/(2w_{0}^2)}%
\]
In our case, this assumption is not based on any profound
insight on the sample, we have chosen this shape because it is
smooth on the top and on the bottom of the grains \cite{Note2} and
because its Fourier Transform is directly evaluated,
\[
FT\left[z_{Pt}\right](k_{x},k_{y})=\frac{1}{2\pi}\int\int
dxdy\;e^{\imath(k_{x}x+k_{y}y)}\;z_{Pt}\left(  x,y\right)
=\frac{h_{0}}{\kappa_{0}^2}\ e^{-(k_{x}^{2}+k_{y}^{2})/(2\kappa_{0}^2)}%
\]
where $k_{i}=1/\lambda_{i}$ is the spatial frequency in each direction
($i =x$ or $y$), and $\kappa_{0}=1/w_{0}$ is the spatial frequency associated
to the width of the grains. For a disordered array of $N$ grains we expect an
``incoherent'' contribution of each grain to the total PSD, and if the grains
cover the surface in a dense arrangement, we expect about one grain in a cell
of width $2w_{0}$, therefore the total number of grains is $N\simeq
(\text{Scan Size}/2w_{0})^{2}=\text{Area}/(4w_{0}^{2})$. The curve $\log\left[
\text{PSD}(1/\lambda)\right]  $ vs $\log(1/\lambda)$ should therefore show a
flat region for small spacial frequencies up to the frequency $1/w_{0}$
corresponding to the width of the grains, and a decrease for higher 
frequencies. Since $e^{-x^{2}}$ decreases faster than any (inverse) power,
this decrease is non-linear, that is, the magnitude of the slope of the
$\log\left[  \text{PSD}(1/\lambda)\right]  $ vs $\log(1/\lambda)$ curve
increases for high frequencies. Correspondingly, this surface is not
self-similar and the notion of fractal dimension is not defined.

This simple model correctly describes the ``good'' PSD curves shown in
\textbf{figures \ref{Fig3}} and \textbf{\ref{Fig4}}. Moreover, with this simple model for the surface morphology
the behaviour of the topographic and amplitude images can be further analysed.
In this context we recall that spatial ($k$) and temporal frequencies ($\nu$) are related
through the scan speed $v$:
\begin{equation}
k=\frac{1}{v}~\nu\label{freqTimeToSpatial}%
\end{equation}
In the graphs shown, the highest and lowest spatial frequencies
(0.04 $\mu$m$^{-1}$ and 20 $\mu$m$^{-1}$) correspond to a temporal
frequency of 1 kHz and 2 Hz.

A characteristic feature of the amplitude PSD curves shown in
\textbf{figure \ref{Fig3}} is that for low spatial frequencies, all
curves have a constant slope $s=2.0\pm0.1$. Since for these spatial
frequencies the PSD curve of the true topography is constant, we
conclude that the slope of the PSD curve is determined by the
filtering properties of the feedback loop. Indeed, if the P/I
controller is modelled by a simple first order electronic circuit
with
characteristic time $\tau_{0}$ we expect transfer functions%
\begin{equation}
g_{\text{topo}}\left(  \nu\right)  =\frac{1}{1+i\tau_{0}\nu}\text{ and
}g_{\text{amp}}\left(  \nu\right)  =1-g_{\text{topo}}\left(  \nu\right)
=\frac{i\tau_{0}\nu}{1+i\tau_{0}\nu} \label{gains}%
\end{equation}
for the topographic and the amplitude signals. Therefore, for low frequencies
the amplitude signal will grow linearly with frequency up to the
characteristic frequency $1/\tau_{0}$. The power of the amplitude signal
increases quadratic with frequency and the $\log\left[\text{PSD}%
[\text{amplitude}(1/\lambda)]\right]  $ vs $\log(1/\lambda)$ curve
should give a straight line with slope $s=2$, as is indeed observed
experimentally.

For this sample with constant PSD of surface roughness up to the characteristic spatial frequency
$\kappa_{0}=1/w_{0}$ the PSD of the amplitude signal is therefore easily
understood taking into account the transfer function of the feedback
loop. A similar analysis for the topographic signal is less evident,
because in the frequency range where the topographic signal is
filtered (for high frequencies) the PSD curve of the true surface
roughness does not follow a simple relation (constant or linear).
Nevertheless, as the P/I\ values are decreased, the topographic
signal is clearly filtered more strongly. Moreover, as the P/I
values are decreased for the different set of images, the
characteristic frequency $1/\tau_{0}$ of the feedback loop also
decreases (spatial frequencies 12.2, 12.2, 8.7, 5, 2.3 nm$^{-1}$ for
P/I 90/45, 45/22.5, 15/7.5, 5/2.5, 1/0.5 respectively in
\textbf{figure \ref{Fig4}}). Finally we note that even though
topographic and amplitude PSD curves are quite different for each
set P/I values, the PSD curves of the difference image give always, within
our experimental error, the same ``master PSD curve''. This is
recognised most easily in \textbf{figure \ref{Fig4}}, where all PSD
curves corresponding to the same kind of data (topography, amplitude
and difference data) have been collected in the same graph in order
to directly visualize how these curves vary as the P/I values are
changed. Very clearly the topographic and amplitude signals vary,
but the difference signal is constant. We stress that what seems to
be a single curve in \textbf{figure \ref{Fig4}} is the
superposition of the five sets of difference data obtained from
topographic and amplitude data shown in \textbf{figure \ref{Fig3}}.

\textbf{Figure \ref{Fig5}} shows a similar data set as that shown in
\textbf{figure \ref{Fig3}}, however in this experiment instead of
varying the P/I values, the scan speed is varied (from left to right:
0.5, 1, 2, 4, 8 lines/s). We first note that as the scan speed is
increased, a characteristic peak moves towards lower frequencies.
This peak is first observed in the second graph \cite{Note1}, and
also in the data of \textbf{figure \ref{Fig3}} at essentially the
same position. We attribute this peak to oscillation of our system
rather than to a true topographic feature. Therefore, according to
relation \eqref{freqTimeToSpatial} as the scan speed is increased,
higher temporal frequencies are measured and the (relative) position
of the peak shifts towards lower values.

As compared to the data shown in \textbf{figure \ref{Fig3}}, only
the first two data sets give ``nice'' PSD\ curves. For high scan
speed, the topography and amplitude images do neither result in
``clean'' PSD curves, nor does the difference data obtained from
each set of images result in a PSD curve that is independent of scan
speed; that is, the difference PSD curves do not lay on a single
``master curve''. In the case of the images shown in \textbf{figure
\ref{Fig5}} we find that data which corresponds to frequencies
higher that the peak shows a behaviour which is not compatible with
the simple first order model of the feedback loop discussed above.
This model essentially predicts a well defined distribution of
topographic and amplitude signal as a function of frequency
according to relation \eqref{gains}. In particular, the PSD curves
of the topographic data acquired at the faster frequencies do not
decrease more strongly than the amplitude data, which should be the
case if the assumption leading to relation \eqref{gains} were strictly valid. Note
that in this region, the amplitude signal is no longer high pass
filtered (the amplitude signal is ``over its maximum'', and this
maximum defines the characteristic frequency
$1/\tau_{0}$ of the feedback loop), which implies that the
topographic data should be high pass filtered. However we observer no high-pass filtering of topography data in
these data sets (compare this region of the PSD curves with
the corresponding behaviour in \textbf{figure \ref{Fig3}}). We
attribute this non-standard behaviour to the fact that at these high
temporal frequencies the SPM setup cannot be considered a simple
(electronic) first order system defined only by the P/I values of
the feedback loop. Instead, also mechanical resonances of the
mechanical SPM setup and possibly even non-linearities of the
tip-sample interaction have to be taken into account, rendering the
tip sample system a much more complicated system in terms of
transfer characteristic. In particular, we believe that for a faithful description of the tip--sample system
orders higher than one, and possibly also non--linearities, have to be taken into account.

Finally, some additional practical issues not discussed so far
should be emphasised. First, we note that for each SFM system the
correct polarity of the error signal will have to be determined
(what is seen low/high by the error signal?). This polarity
determines the sign of relation \eqref{ZtrueSumOrDiff}, that is,
whether the error signal has to be added or subtracted (as assumed
in the present work) in order to obtain the ``true'' topography. A
second issue not discussed yet is the correct setpoint for image
adquisition. In order for the error signal to faithfully reproduce
the surface morphology, it has to be a linear function of tip-sample
distance. Therefore the setpoint has to be chosen so that there is
``enough signal'' when passing over high and low surface features.
This is illustrated in \textbf{figure \ref{Fig6}} for the case of
the oscillation amplitude as error signal. If possible, the setpoint
$i_{0}$ of the interaction should be chosen such that the values
between $i_{0}-\delta z_{rms}$ and $i_{0}+\delta z_{rms}$ depend linearly
on the tip-sample distance (recall that the error signal is calibrated in length units), where $\delta
z_{rms}$ is the rms roughness of surface morphology.  If the setpoint $i_{0}$ is chosen too
close either to the free oscillation amplitude ($a^b_{\text{set}}=$
bad setpoint in \textbf{figure \ref{Fig6}}) or to the minimum
amplitude needed to sustain a stable oscillation, then a small
surface roughness will move the tip-sample system from the linear
part of the amplitude vs distance curve. If this is the case, the
error signal will not any more contain the correct information about
the surface morphology, and the true surface morphology cannot be
reconstructed as discussed in the present work.

\section{Discussion}

We have shown that the correct determination of surface
morphology using SPM techniques is extremely demanding and may be easily affected
by experimental parameters. The description of surface morphology using PSD
curves is a very powerful tool, but requires very good experimental raw data.
In order to exploit the full potential of PSD analysis every experimental data
point has to be a faithful representation of the true surface. For feedback
based and sequential imaging techniques such as SPM, this is an very
difficult task. The data shown here proves that in most cases if only
topography data is acquired, the measured morphology is significantly affected
by the feedback response of the SPM system. Then, the PSD curves
calculated from this experimental data do not correspond to that of the true
topography. Instead, either features are ``lost'' due to low pass filtering or
features are ``created'' due to oscillation of the feedback loop. In most
cases the PSD curves obtained from topographic images depend strongly on the
parameters used for data adquisition (scan speed, P/I values of the feedback
loop), and it is not clear which, if any, is the ``good'' curve. When the
error signal of the feedback loop is acquired and analysed, the characteristic
response time of the feedback loop can be determined. This response time
determines, together with the scan speed, the maximum spatial frequency up to
which the topographic data is measured faithfully. In addition, possible
oscillation of the feedback loop can also be recognised in the error signal.
The error signal can thus be used to control the quality of the topographic
image.

If the error signal is correctly calibrated (in length units: [nm]),
then the topographic and the error signal can be summed (with the correct
sign!) in order to give a ``true'' image that is a faithful representation of
the surface morphology. In particular, this combined image does not depend on
the particular set of parameters used for image adquisition. Interestingly, we
observe that the best data is not acquired with high feedback parameters,
since these result in oscillation of the feedback loop; imperceptible in the
topographic and error signal data, but clearly observed in the corresponding
PSD curves. Therefore for faithful imaging of the surface, the feedback parameters
should not be pushed too high, instead, as discussed in this work, the
calibrated error signal should be used to ``recover'' the small scale surface
morphology, which is (low-pass) filtered by the feedback loop in the
topographic image.

Since the nanoscale surface morphology determines many surface related
processes such as friction, adhesion, wetting as well as many others, its
correct determination is a fundamental issue in nanoscience. From a
theoretical point of view, the PSD of surface topography is a basic tool to
describe the statistical properties of surfaces and is used as a key parameter
for the description of surface morphology in modern theories of friction and
adhesion. Accordingly, its precise experimental measurement as proposed in
this work is a fundamental issue for nanoscience, and we strongly believe that
the approach presented in this work substantially improves the performance of
any Scanning Force Microscope when using contact mode SFM as well as AM-DSFM,
which are the two modes used for most imaging applications.

\section{Acknowledgments}

This work was supported by the Ministerio de Ciencia e
Innovaci\'{o}n (MICINN, Spain) through the proyects
``ForceForFuture'' (CONSOLIDER programme, CSD2010-00024)  and
``CONAMA-Nano'' (MAT2010-21267-C02-01), as well as by the Comunidad
Aut\'{o}noma de la Regi\'{o}n de Murcia  through the proyect
``C\'{e}lulas solares org\'{a}nicas de la estructura molecular y
nanom\'{e}trica a dis\-po\-si\-ti\-vos operativos
macrosc\'{o}picos'', Plan de Ciencia y Tecnolog\'{i}a de la
Regi\'{o}n de Murcia 2007-2010. In addition, INC also acknowledges
support from the Comunidad Aut\'{o}noma de la Regi\'{o}n de Murcia
through a scholarship from the Fundaci\'{o}n S\'{e}neca.

\bibliographystyle{Biblio}
\bibliography{PsdErrorAndTopo}

\clearpage
\newpage

\section{Figure Captions}

\subsection{Figure 1}

\begin{figure}[!ht]
\centering
\includegraphics[width=15cm]{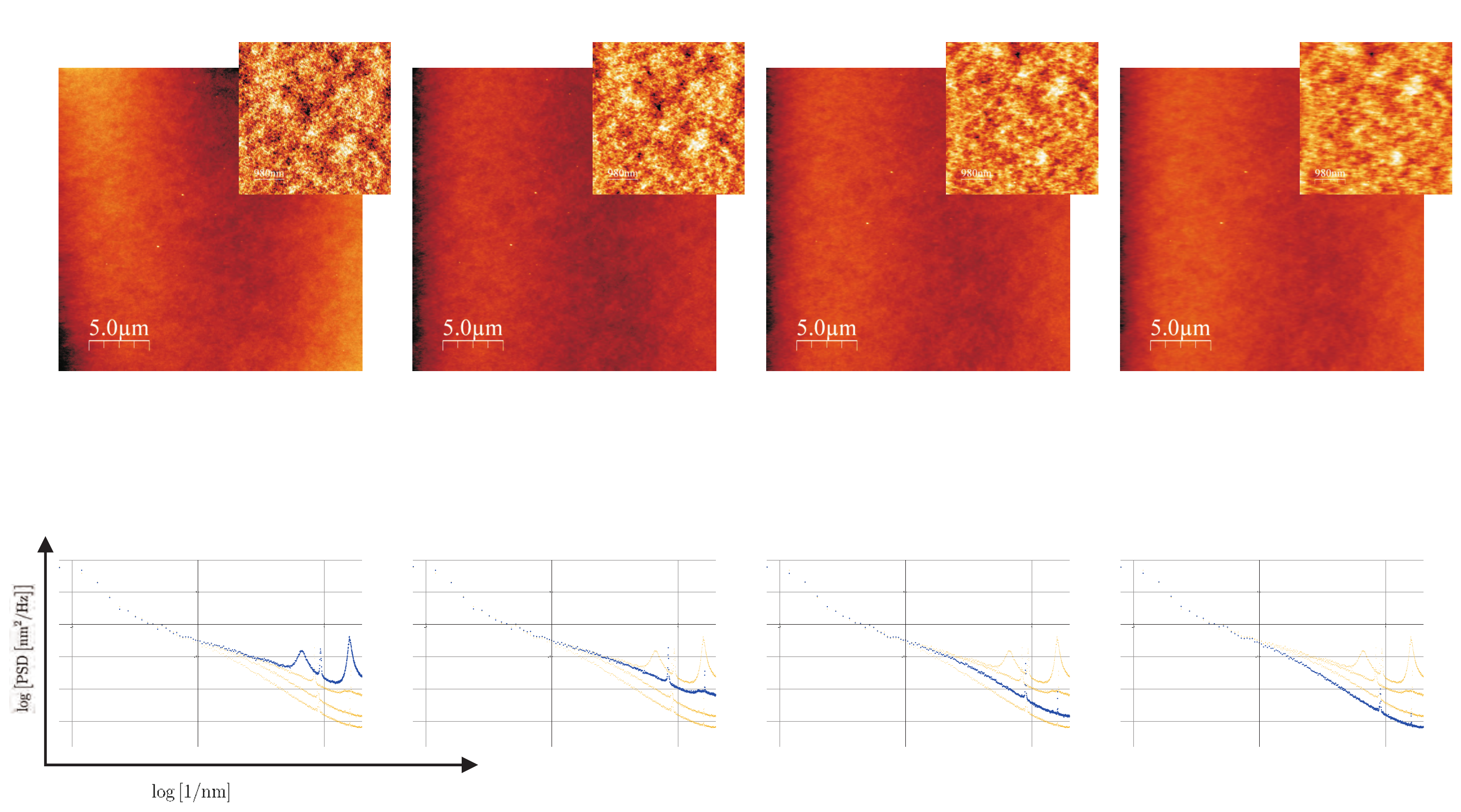}
\caption{}\label{Fig1}
\end{figure}

Top row: Topography images of the surface of a commercial glass
cover slide acquired at a scan frequency of 1 line/s, but at
different settings for the feedback loop. Feedback response was
decreased from left to right: proportional/integral parameters are
125/25, 80/16, 40/8 and 20/4 (in arbitrary units). Image size is
$1024\times1024$ and the total acquisition time was about 18
minutes. The insets shows enlarged regions of each topographic image
with an amplified grey scale. Lower row: graphs of the PSD curves of
surface roughness calculated from the corresponding (large scale)
topographic image. Lateral size of the larger images is 25 $\mu$m,
smaller images show a zoom of 5 $\mu$m. The total grey scale of all
large scale images is 8 nm and 1 nm for all insets. Bottom
row: PSD curves calculated from the topographic images; the
graphs show the logarithm of the PSD of surface roughness plotted versus
the logarithm  of the inverse length scale. For each graph the
thinner yellow lines show the PSD curves of all the other images, while the
thicker blue line shows the PSD curve for the correspondent topographic
image shown in the same column. The grid lines for the PSD graphs
are $\Delta \log[\kappa]=1$ (horizontal axis) and $\Delta
\log[\text{PSD}]=1$ (vertical axis).

\clearpage
\newpage

\subsection{Figure 2}

\begin{figure}[!ht]
\centering
\includegraphics[width=15cm]{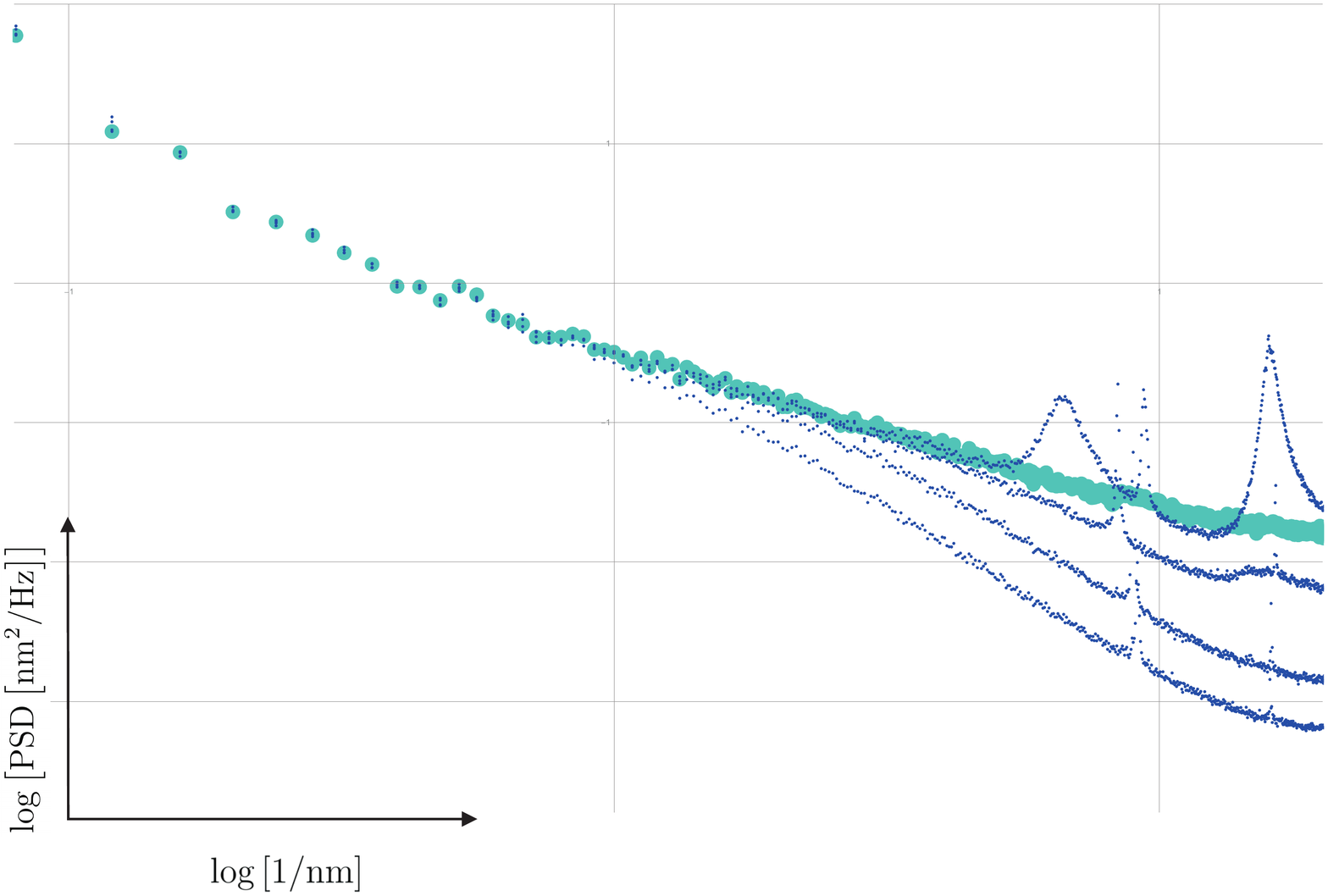}
\caption{}\label{Fig2}
\end{figure}

PSD curves calculated from the topographic images shown in
\textbf{figure \ref{Fig1}} as well as from the discussion above.The
graphs show the logarithm of the PSD of surface roughness plotted
versus the logarithm of the inverse length scale. The thinner lines
correspond to the PSD curves of the individual images shown in
\textbf{figure \ref{Fig1}}, the thicker line to a master curve
described in the main text. The grid lines for the PSD graphs are
$\Delta \log[\kappa]=1$ (horizontal axis) and $\Delta
\log[\text{PSD}]=1$ (vertical axis).

\subsection{Figure 3}

\begin{figure}[!ht]
\centering
\includegraphics[width=17.2cm]{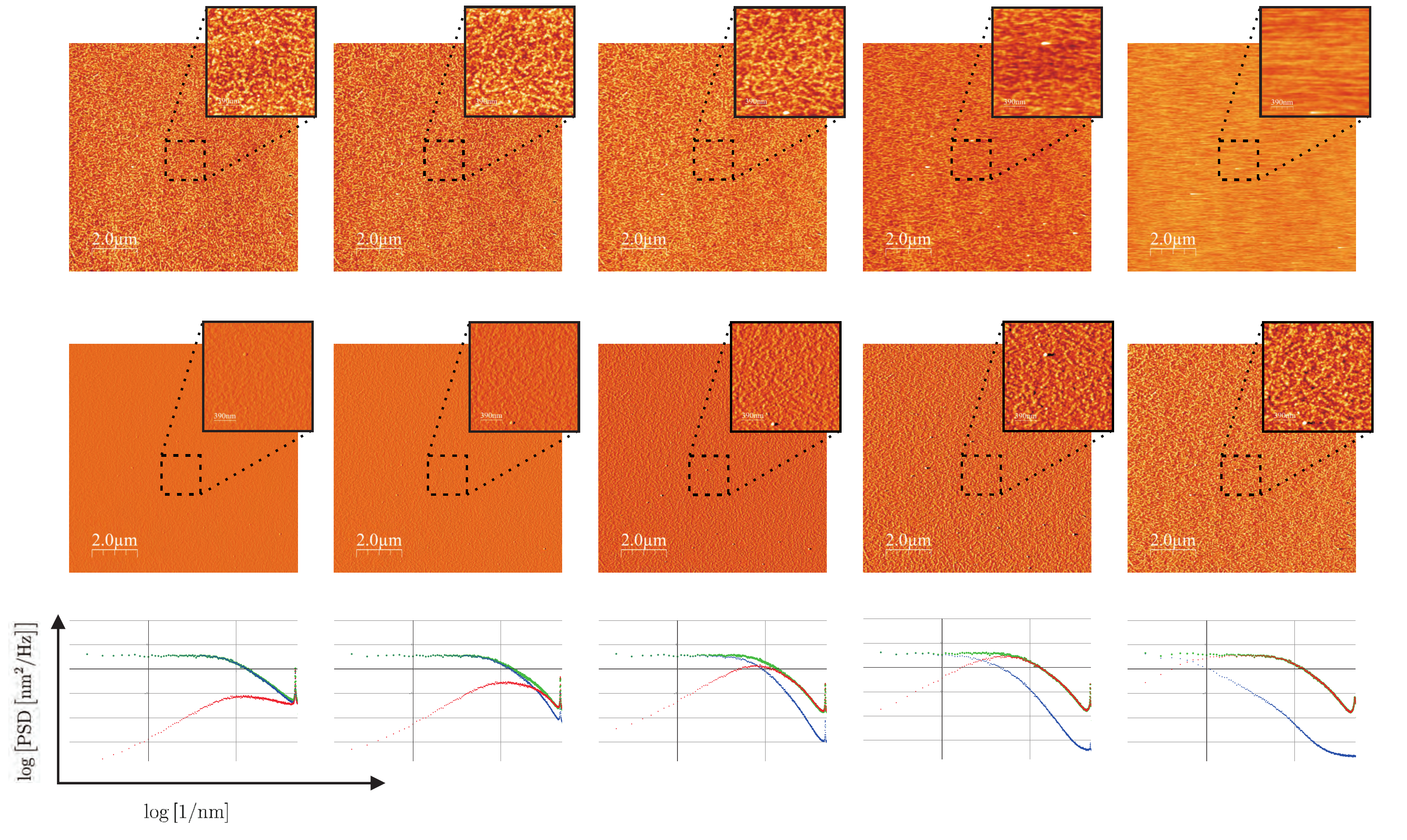}
\caption{}\label{Fig3}
\end{figure}

Images of a Platinum surface taken all at a scan frequency of 1
line/s, but at different settings for the feedback loop (from left
to right: 90/45, 45/22.5, 15/7.5, 5/2.5, 1/0.5). The images and
graphs in the same column correspond to a common data set, since the
corresponding images have been acquired simultaneously at a fixed
values of the feedback loop. The upper row shows topographic images,
the middle row amplitude images and the lower row graphs of the PSD
curves of surface roughness. In each graph, the PSD of surface
roughness has been calculated for the topography and amplitude image
shown in the corresponding column, as well as for the difference
$z_{\text{true}}=z_{\text{fb}}-\Delta I_{\text{err}}$ (corresponding
image not shown). Lateral size of the larger images is 10 $\mu$m,
smaller images show a zoom of 1 $\mu$m. The total grey scale of all
images is 5 nm (large and small scale images as well as topographic
and amplitude images). Lower row: PSD curves calculated from the
topography and amplitude images as well as the difference data (image not shown).
The grid lines for the PSD graphs
are $\Delta \log[\kappa]=1$ (horizontal axis) and $\Delta
\log[\text{PSD}]=1$ (vertical axis). Green lines correspond to the PSD curves of topography plus error signal,
blue curves to the topography and red curves to the error signal.

\subsection{Figure 4}

\begin{figure}[!ht]
\centering
\includegraphics[width=17.2cm]{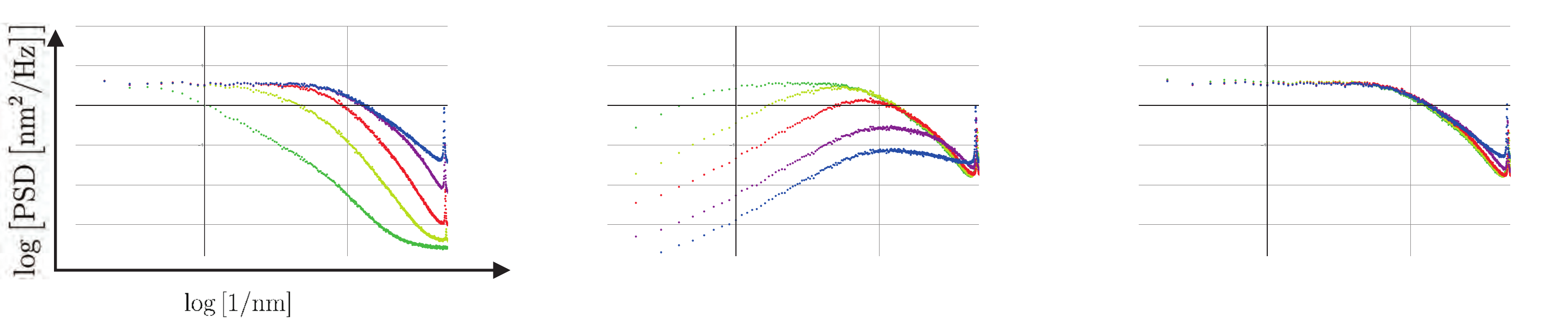}
\caption{}\label{Fig4}
\end{figure}

PSD curves corresponding to the topography and amplitude images
shown in \textbf{figure \ref{Fig3}} as well as of the difference
data. Each graph shows the $\log\left(\text{PSD}[1/\lambda]\right)$ vs
$\log(1/\lambda)$ curve for the data acquired at different P/I
values. While the different curves are clearly distinguished in the
topography (a) and amplitude (b) PSDs, the curves corresponding to
the difference (c) essentially fall on the same ``master curve''.
The grid lines for the PSD graphs
are $\Delta \log[\kappa]=1$ (horizontal axis) and $\Delta
\log[\text{PSD}]=1$ (vertical axis).

\subsection{Figure 5}

\begin{figure}[!ht]
\centering
\includegraphics[width=17.2cm]{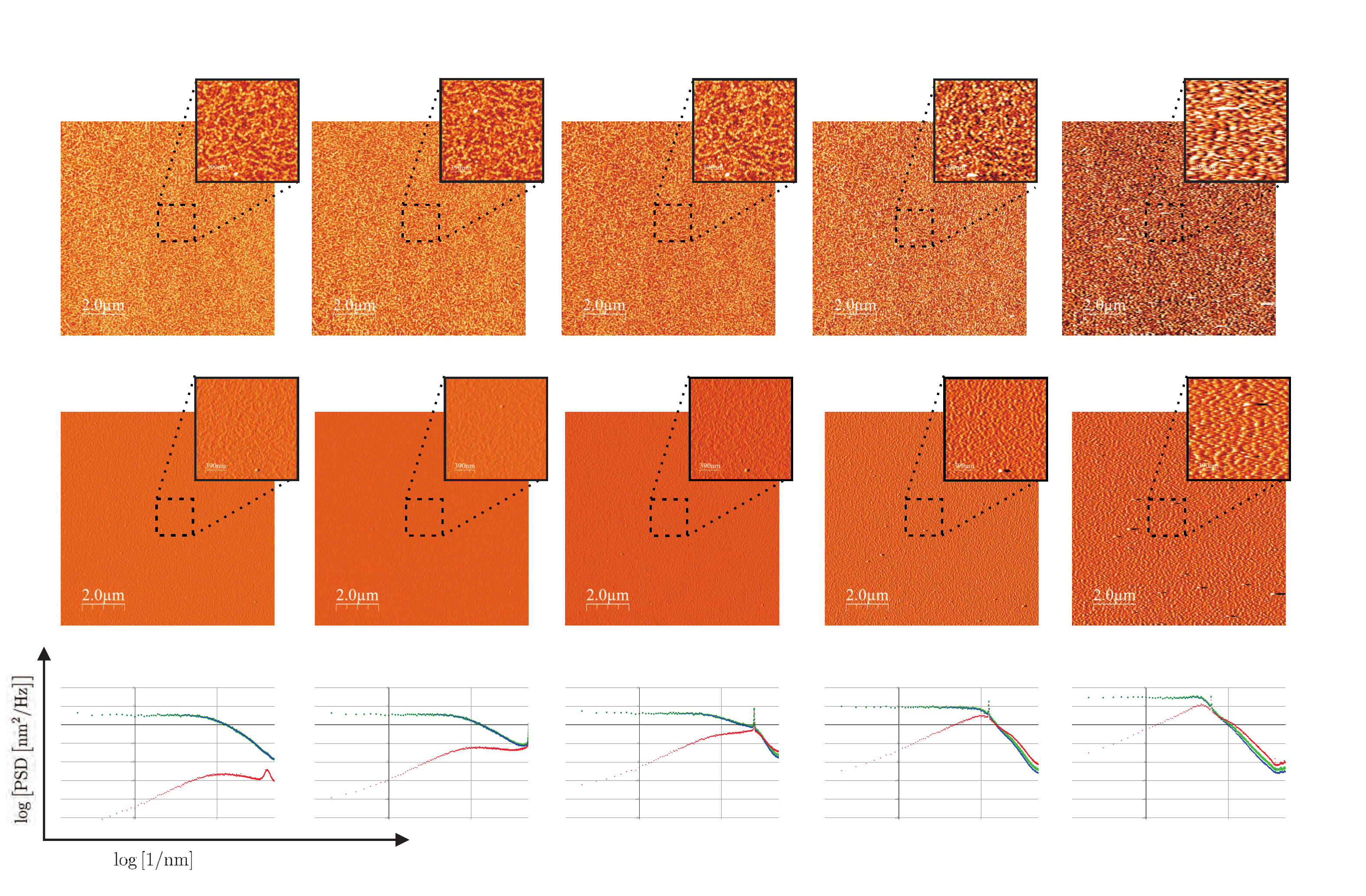}
\caption{}\label{Fig5}
\end{figure}

Images of the same Platinum film as shown in \textbf{figure
\ref{Fig3}}. The images shown were acquired at the same value of the
feedback loop, but at different imaging speeds (from left to right:
0.5, 1, 2, 4, 8 lines/s). As previously the upper row shows
topographic images, the middle row amplitude images and the lower
row graphs of the corresponding PSD curves of surface roughness.
Again, lateral size of the larger images is 10 $\mu$m, smaller
images show a zoom of 1 $\mu$m. The total grey scale of all images
is 5 nm. The grid lines for the PSD graphs are $\Delta
\log[\kappa]=1$ (horizontal axis) and $\Delta \log[\text{PSD}]=1$
(vertical axis).

\subsection{Figure 6}

\begin{figure}[!ht]
\centering
\includegraphics[width=12cm]{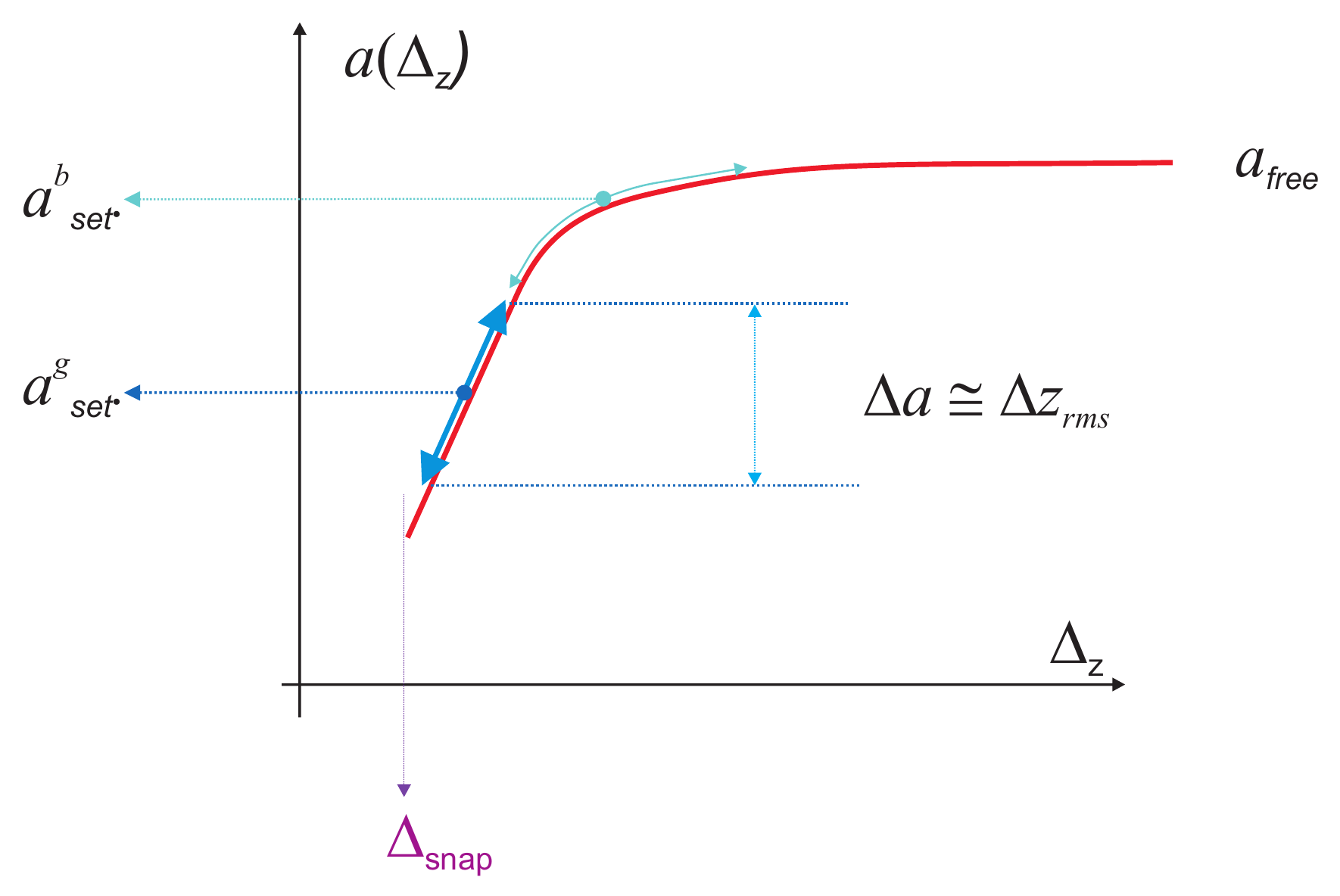}
\caption{}\label{Fig6}
\end{figure}

Schematic representation of the oscillation amplitude vs the
tip-sample distance. For large tip sample distances, the free
oscillation amplitude is measured. As the oscillating tip interacts
with the surface the oscillation decreases linearly with tip sample
distance. For small oscillation amplitude, the energy pumped into
the cantilever by the external driving circuit is not sufficient to
compensate for the losses induced by the tip-sample interaction, and
the oscillation stops. If the feedback loop does not respond
instantaneously to height variations as the tip is scanned over the
surface, height variations will result in variations of the
oscillation amplitude. For the kind of applications proposed in this
work, the tip-sample system has to stay in the linear regime of the
amplitude vs distance curve.

\end{document}